\documentclass[12pt,a4paper]{article}
\usepackage{amsmath,amssymb,graphicx,epsfig,cite,color}

\usepackage[left=2.7cm,right=2.7cm,top=2.5cm,bottom=2.5cm]{geometry}
\usepackage{subfigure}
\usepackage[T1]{fontenc}
\usepackage{multirow}
\usepackage[table]{xcolor}
\usepackage{booktabs}

\usepackage{xcolor,colortbl}

\usepackage{graphicx}
\usepackage{lscape}
\usepackage{appendix}
\usepackage{multirow}
\definecolor{DarkPastelGreen}{rgb}{0.01,0.69,0.28}
\definecolor{amethyst}{rgb}{0.6, 0.4, 0.8}
\definecolor{darkcyan}{rgb}{0.0, 0.55, 0.55}
\usepackage[colorlinks=true,linkcolor=amethyst,citecolor=darkcyan,urlcolor  = DarkPastelGreen]{hyperref}%
\allowdisplaybreaks[1]

\begin{document}




\title{
         {\Large
                 {\bf
\color{darkcyan}{ CGAN-Based Framework for Meson Mass and Width Prediction}
                 }
         }
      }

\author{\vspace{1cm}\\
	{\small
	S.~Rostami$^{1}$,
	M.~Malekhosseini$^{2}$,
		M.~ Rahavi Ezabadi$^{2}$,
		K. Azizi$^{2,3}$
		\thanks{Corresponding author, e-mail: kazem.azizi@ut.ac.ir}
		}
\\
  {\small$^1$  School of Physics, Institute for Research in Fundamental Sciences (IPM),}\\
			{\small  P. O. Box 19395-5531, Tehran, Iran}\\
	{\small $^2$ Department of Physics, University of Tehran, North Karegar Avenue, Tehran 14395-547, Iran}\\	
	{\small $^3$ Department of Physics, Dogus University, Dudullu-\"Umraniye, 34775 Istanbul, T\"urkiye}\\
	} 

\date{}

\begin{titlepage}
\maketitle
\thispagestyle{empty}

\begin{abstract}
Mesons play a crucial role in understanding the strong interaction in the 
framework of quantum chromodynamics (QCD).  However, the mass and 
decay width of several ordinary and exotic mesons remain experimentally 
undetermined.  In this work, we propose a novel application of advanced 
machine learning techniques to deal with this challenge. 
Due to the limited available meson datasets, 
traditional data-driven methods are norm
To overcome this, we employ a Conditional Generative Adversarial 
Network (CGAN) to generate synthetic meson data based on known physical parameters. 
This not only augments the dataset but also retain the underlying physics 
of the original mesons data. 
With the extended dataset, we train multiple copies of CGAN and apply a 
bagging technique to predict uncertainties, improving  the robustness and 
reliability of the predictions. As our findings indicate, the CGAN models are 
capable of well describing meson properties and their structure relations,
offering a potent novel instrument for hadron spectroscopy.
This calculation opens a promising future for data-driven hadron physics studies.
\end{abstract}
 
\end{titlepage}
\section{Introduction}

Particle physics investigates the fundamental particles and forces that constitute the building blocks of the universe. Within this broad field, hadron physics focuses specifically on the properties of  hadrons,  composite particles made of quarks and gluons. 
The interactions and properties of the hadrons, including mesons and baryons, are crucial to understanding the strong force, 
as they represent the physical manifestation of quarks and gluons bound together by this force. 

Quantum Chromodynamics (QCD) is the theoretical framework that describes the interactions between the quarks and gluons, the fundamental constituents of hadrons. QCD explains how the strong force operates at the quark level, 
dictating their binding within hadrons and giving rise to phenomena like confinement, where quarks are never found in isolation \cite{Gell-Mann:1964ewy,Leutwyler:2012ax,Yukawa:1935xg,ParticleDataGroup:1986kuw,ParticleDataGroup:2024cfk,Goldstone:1962es,Field:1989uq,Ellis:1996mzs,Gross:2022hyw}.

QCD has been extremely successful in explaining the strong force 
and the hadronic interactions. Also, its experimental verification is 
one of the major triumphs of modern physics \cite{Ellis:1996mzs,Gross:2022hyw,Rabbertz:2013vxa,ATLAS:2008xda,CMS:2008xjf,Collins:1981uk,ZEUS:1993ppj,Harriman:1990hi}. 
However, many aspects of QCD, especially in the non-perturbative regime, 
continue to be an area of active research \cite{Ladinsky:1993zn,Barnett:1976ak,Altmann:2024icx}. 

The low-energy behavior 
of mesons, such as their masses and decay properties, is a window into the 
non-perturbative aspects of QCD. 
For instance, lattice QCD is a computational technique that uses a grid-like 
	framework to simulate the strong interactions between quarks and gluons. 
	This approach help physicists to study complex, 
	non-perturbative behavior of the mesons from first principles, providing valuable insights into their masses and decay properties \cite{Cichy:2018mum, Yang:2014xsa}.

In experiments, measurements of mesons 
mass and decay width help to refine our understanding of how quarks and 
gluons interact when they are bound together to form mesons and other hadrons.
So it provides essential input to test theoretical models of QCD in this low-energy regime \cite{Lattes:1947mx,Lattes:1947my,Godfrey:2015dia,D0:2007vzd,CMS:2021mzx,PHENIX:2015cea,ALICE:2018lyv,CMS:2018eso,Pevsner:1961pa,Gell-Mann:1962hpq,Gell-Mann:1961jim,Barbaro-Galtieri:1969kpv,Crennell:1968zza}.

Theoretical models in  hadron physics are crucial for interpreting experimental results, especially 
those derived from high-energy collisions at facilities such as the Large Hadron Collider (LHC) \cite{Crennell:1970uq,Micu:1968mk,Nambu:1957wzj,Chew:1960iv,Jacob:1961zz,Godfrey:1985xj,Celmaster:1977vh,DeRujula:1975qlm,Colangelo:2004vu,Lahde:1999ih,Ebert:1997nk,Bardeen:2003kt,Ni:2021pce,Guo:2006rp,ALICE:2024bhk,Okubo:1963fa}.
While there has been significant experimental and theoretical progress in 
the hadron physics, particularly regarding exotic states, the internal structure 
and quark-gluon configurations of some ordinary and exotic hadrons remain unclear. 
Additionally, the mass and decay widths of several hadrons, including both ordinary 
and exotic mesons, have yet to be precisely determined \cite{Ali:2017jda,Bhattacharya:2024neb,Nefediev:2024ihz,Song:2024ngu,Farina:2023oqk,Wang:2023ivd,Guo:2023xyf,Liu:2023hhl,Wang:2023whb}. 

At this stage,  modern simulation techniques have become crucial. As the LHC generates 
vast amounts of data, simulations are essential tools for unraveling the complexities 
of hadronic interactions \cite{Stelzer:1994ta,Alwall:2014hca,Hsieh:1991ti,Kublbeck:1990xc,Montvay:1975bf,Jamin:1991dp,Sjostrand:2006za,deFavereau:2013fsa,Allison:2006ve}.
 This dynamic interplay between theory, experiment, 
and simulation offers a unique opportunity to advance our understanding of hadronic 
interactions and the complex dynamics of quarks and gluons, particularly in 
the non-perturbative regime, where traditional analytical methods are less effective in 
determining the mass and decay widths of hadrons. 

Traditional methods, such as Monte 
Carlo simulations, often require significant computational resources and time, especially 
when dealing with complex systems or high-precision calculations. While advances in 
technology and new computational techniques, such as machine learning (ML) \cite{Schwartz:2021ftp,Butter:2022rso,Dubinski:2023fsy}, 
are helping to alleviate some of these challenges, traditional methods can still be quite demanding. 

The ML algorithms, particularly those based on deep learning-based generative models, 
can produce realistic synthetic data that mimics experimental outcomes. This not 
only enhances the accuracy of simulations but also allows for the exploration of parameter 
spaces that may not be feasible with traditional methods \cite{Kansal:2022spb,Hashemi:2019fkn,Salt:2024jkn,Ahmad:2024dql,Kita:2024nnw}. By learning patterns from 
large datasets, ML models can identify relationships between complex variables, 
optimize simulation parameters, and even predict outcomes that would typically 
require time-consuming calculations. For example, ML algorithms can be used to 
accelerate the process of hadronization, predict hadron spectra, or automate the 
identification of particle decay modes. Additionally, ML can be employed to analyze 
and interpret experimental data more effectively, allowing for faster and more precise 
extraction of physical quantities such as cross-sections, particle trajectories, and event 
classification. These advancements are helping to bridge the gap between traditional 
simulation methods and the vast amounts of data generated by modern high-energy 
physics experiments, making it possible to explore new regions of parameter 
space and extract insights more efficiently \cite{deOliveira:2017pjk,Radovic:2018dip,Albertsson:2018maf,Guest:2018yhq,Lonnblad:1990qp,Malekhosseini:2024eot,Roe:2004na,Mamaev:1977tx,Edelen:2016dqu,Gal:2020dyc,Bahtiyar:2022une}. 

This collaboration between 
the simulation and the ML is paving the way for deeper insights into the hadron physics. 
More accurate event classification, improving parameter estimation, and facilitating the 
development of sophisticated simulations in hadronic data are advanced techniques that 
enrich our understanding of the hadron physics \cite{Guest:2018yhq,Lonnblad:1990qp,Ghosh:2022zdz,Guest:2016iqz,ATLAS:2017gpy,Louppe:2017ipp,Baldi:2016fzo}. 

Deep learning-based 
generative models are able to generate new data samples from learned distributions. 
They have gained significant attention due to their ability to produce high quality outputs \cite{Kita:2024nnw,Chan:2023ume,Otten:2019hhl,Paganini:2017hrr,Bedkowski:2024ilr}. 
Common methods for deep generative models include variational autoencoders (VAEs) \cite{Dohi:2020eda,Doersch:2016eco}, 
normalizing flows (Nfs) \cite{Durkan:2019nsq,Quetant:2024ftg}, and generative adversarial networks (GANs) \cite{deOliveira:2017pjk,Goodfellow:2014upx,DiSipio:2019imz}. 
For instance, the VAE framework has been introduced to generate realistic and diverse 
HEP events. This model benefits from several techniques in the VAE 
literature to simulate high-fidelity jet images \cite{Dohi:2020eda}. Normalizing flows is one of the approaches 
employed to directly generate full events at the detector level from Parton-level information. 
As such, this research represents an important step in advancing generative modeling techniques 
in high-energy physics \cite{Quetant:2024ftg}.

 GANs are a class of deep learning 
generative models where two neural networks, a generator, and a discriminator, compete against 
each other to produce realistic synthetic data. The generator creates new data samples, while the 
discriminator evaluates them against real data, guiding the generator to improve its output \cite{Goodfellow:2014upx}. 

In recent years, GANs have become influential techniques in a variety of scientific fields,
including particle physics. GANs have been effectively used to generate 
high-fidelity event samples in collider physics, enabling the production of complex multi-particle 
final states that closely resemble actual collision data. This application helps in simulating and 
analyzing particle interactions more efficiently \cite{deOliveira:2017pjk,Chan:2023ume,DiSipio:2019imz,Matchev:2020tbw,Simsek:2024zhj,Paganini:2017dwg}. 
The hadronization plays a crucial role in simulating high-energy experiments. 
Ref. \cite{Chan:2023ume}, has introduced a protocol for training a 
deep generative model for hadronization, employing a GAN framework with a permutation-invariant 
discriminator. The authors assert that their work marks a significant advancement in the ongoing effort 
to develop, train, and incorporate the ML-based models of hadronization into parton shower Monte Carlo 
simulations.

 A specialized variant of GANs, known as Conditional GANs (CGANs), modify the GAN setup 
by conditioning both the generator and the discriminator on additional information. This could be labels, 
images, or any other type of data that specifies a desired output, helping generate more targeted data \cite{Mirza:2014dfp}. 
In the hadron physics, CGANs can be used for the event generation, the simulation of HEP collisions, 
the parameter estimation and identifying anomalies in new data, such as potential signals of 
new physics \cite{Simsek:2024zhj,Pearce-Casey:2023gst}. 
A recent study has shifted its focus from traditional full-simulation 
methods to investigate the potential of a deep learning-based CGAN. The research presents 
a fast simulation technique that uses CGANs to convert calorimeter images, offering the 
potential to significantly reduce both computational time and disk space requirements 
for the LHC and future high-energy physics experiments \cite{Simsek:2024zhj}. 

The generative models, 
due to their ability to create synthetic data, have become increasingly valuable in the 
field of generative data augmentation \cite{Matchev:2020tbw,Chang:2021jne}. The data augmentation techniques are 
commonly employed when the available data for analysis or simulation is limited, as 
this limitation can lead to reduced model accuracy and generalization. By artificially 
increasing the diversity of the training data, data augmentation helps improve the 
robustness of models, especially in fields like  ML and HEP, where acquiring large 
amounts of labeled data can be costly or time-consuming. Despite significant achievements 
in the HEP, including the cataloging of numerous mesonic and baryonic states by the 
PDG \cite{ParticleDataGroup:1986kuw,ParticleDataGroup:2024cfk}, the available dataset for studying these states through the deep generative 
models remains limited. 

In this context, data augmentation techniques can provide a useful 
solution \cite{Matchev:2020tbw,Chang:2021jne,Dubost,Karras}. It should be noted that two methods have been presented to augment 
the available hadronic data so far. In the first method, experimental mass errors are added to 
and subtracted from their central values, while the quantum numbers remain fixed. 
This results in the training data being resampled twice. The second method employs 
Gaussian noise resampling, using a Gaussian probability density function. In this approach, 
random data points are generated based on the mean values and errors from the 
training dataset, with the quantum numbers of the hadrons held constant. 
The hadronic data undergoes up to 9 data replications using this method \cite{Bahtiyar:2022une,Bahtiyar:2022wph}. 
Building on this foundation, we have developed, for the first time, a CGAN model 
specifically designed to augment existing meson data. By harnessing the power of 
CGANs, we generate synthetic meson data that closely mirrors the distribution and 
characteristics of real-world measurements. This approach has the potential to 
advance data-driven studies in hadron physics, providing a more comprehensive 
dataset for further analysis, model training, and improved predictions of mesonic 
properties. 

The machine learning approach does have some advantages, including the ability to process complex, non-linear correlations in hadronic data efficiently, decreased computational expenses compared to traditional lattice or Monte Carlo simulations, and exploring parameter spaces otherwise unreachable. It also facilitates data augmentation in cases of limited experimental measurements to augment model robustness and predictive power. However, the method is not without its shortcomings: it depends on the quality and size of the training dataset, has the potential to inflict model assumption or preprocessing bias, and has no interpretability in comparison to direct first-principles approaches such as lattice QCD. Therefore, our outcomes have to be considered supplementary rather than substitutive to established experimental and theoretical methods.

One key controversy in the field involves the quark content of certain 
mesons, with ongoing ambiguity about whether they should be classified as 
ordinary mesons or exotic states, such as tetraquarks. Additionally, the 
masses and decay widths of both ordinary and exotic mesons remain 
poorly measured, contributing to the uncertainties and debates within 
hadron physics. Notably, the fundamental properties of the meson 
spectrum have been used to estimate  the masses of baryons, pentaquarks, 
and other exotic hadrons \cite{Gal:2020dyc}. Inspired by this framework, we developed 
deep neural networks (DNNs) to more accurately estimate the mass and 
decay width of both ordinary and exotic mesons \cite{Malekhosseini:2024eot}. 

In this work, we first employ a CGAN model to generate synthetic meson data that retain the key physical properties of the original mesons. 
By augmenting the limited experimental dataset of mesons with this synthetic data. We then average predictions from a set of CGAN models through bagging so that we are able to estimate uncertainties and give better estimates for both ordinary and exotic mesons.  It allows us to make more accurate and reliable predictions.
Our results agree with experimental data available and provide an useful tool for studying mesons not easy to analyze.
As a whole, the present contribution opens up the possibility of employing sophisticated machine learning methods in hadronic physics studies.

\section{Generative Adversarial Networks}\label{GAN}

Neural networks (NNs) are computational models based on the architecture and processes of the human brain.
They are designed to identify patterns and relationships within data through a network of interconnected layers. 
Each layer comprises artificial neurons, which perform fundamental computations to process and transform information:
\begin{itemize}
    \item \textbf{Input Layer:} This layer takes in raw data features and transmits them to the subsequent layers, with the number of neurons matching the number of input features.
    \item \textbf{Hidden Layers:} These layers process inputs using weights and biases, applying activation functions to capture non-linear relationships. The number and size of hidden layers determine the model's capacity.
    \item \textbf{Output Layer:} This layer generates the model's final predictions, 
    translating the processed information from the hidden layers into a usable result.
    The number of neurons here corresponds to the number of outputs required (e.g., for a classification task, it could be the number of classes).
\end{itemize}

Each neuron computes a weighted sum of its inputs, adds a bias term, and applies an activation function, as described by the following equation:
\begin{equation}
\text{Output} = f\left(\sum_{i=1}^{n} w_i x_i + b\right),
\end{equation}
where $x_i$ are the inputs, $w_i$ are the weights, $b$ is the bias, and $f$ is the activation function.

Common activation functions include:
\begin{itemize}
    \item \textbf{Sigmoid:} $f(x) = \frac{1}{1 + e^{-x}}$, maps inputs to a range between 0 and 1.
    \item \textbf{ReLU (Rectified Linear Unit):} $f(x) = \max(0, x)$, introduces sparsity and mitigates the vanishing gradient problem.
    \item \textbf{Tanh:} $f(x) = \tanh(x)$, maps inputs to a range between -1 and 1.
\end{itemize}

The training process involves minimizing a loss function, 
which quantifies the difference between predicted and actual outputs. 
This is achieved using optimization algorithms such as gradient descent, 
which iteratively adjusts the weights and biases to reduce the loss. The update rule is:
\begin{equation}
\theta \rightarrow \theta - \eta \frac{\partial \mathcal{L}}{\partial \theta},
\end{equation}
where $\theta$ represents the model parameters (weights and biases), $\eta$ is the learning rate, and $\mathcal{L}$ is the loss function.

Backpropagation efficiently computes gradients by propagating errors from the 
output layer back through the earlier layers. This process ensures that each layer 
adjusts to minimize its contribution to the overall error.

NNs, with their layered architecture, activation functions, 
and loss functions, have revolutionized numerous domains by learning 
complex patterns from data. However, traditional  NNs  are 
primarily designed for predictive tasks, such as classification or regression, 
which involve mapping inputs to outputs. To address the challenge of data 
generation and expand the capabilities of NNs, researchers have 
introduced advanced architectures like GANs. 
GANs build on the foundational principles of NNs but take them 
further by incorporating two competing networks -generator and discriminator-
that work together to create data indistinguishable from the real dataset.

GANs are a class of ML frameworks
introduced by Ian Goodfellow and his colleagues  in 2014 \cite{Goodfellow:2014upx}. 
GANs comprise two NNs, a generator ($G$) and a discriminator ($D$), 
which are trained simultaneously through adversarial learning. 
The generator aims to create realistic data samples, while the 
discriminator's task is to differentiate
between real and generated data.
The adversarial nature of this process enables GANs to generate synthetic data that closely corresponds to real data.

The training process of GANs is formulated as a min-max optimization problem, 
where the generator and discriminator engage in a two-player game.
 The objective function is given by:
\begin{equation}
\min_G \max_D \mathcal{L}(D, G) = \mathbb{E}_{x \sim p_{\text{data}}(x)}[\log D(x)] + \mathbb{E}_{z \sim p_z(z)}[\log(1 - D(G(z)))],
\end{equation}
where $p_{\text{data}}(x)$ represents the distribution of real data, $p_z(z)$ is the prior distribution of the input noise $z$, $G(z)$ generates fake data, and $D(x)$ outputs the probability that $x$ is real.

The adversarial learning framework ensures that the generator improves over time by 
"fooling" the discriminator, while the discriminator simultaneously becomes 
more proficient at distinguishing real data from generated data.
This adversarial process continues until the generator produces data that 
the discriminator can no longer consistently differentiate from real data.

Despite their success, GANs face several challenges, including 
mode collapse, where the generator produces a limited range of outputs, 
and training instability arising from the min-max optimization process.
Various techniques have been proposed to address these issues, 
such as Wasserstein GAN (WGAN) and gradient penalty, 
which enhance training stability by modifying the loss function.

CGANs build upon the GAN framework by integrating extra information, 
such as class labels  $y$ or specific features, 
into the data generation process. 
In CGANs, the generator and discriminator both rely on conditioning with $y$, 
and the objective function is modified accordingly
\begin{equation}
\min_G \max_D \mathcal{L}(D, G) = \mathbb{E}_{x, y \sim p_{\text{data}}(x, y)}[\log D(x|y)] + \mathbb{E}_{z \sim p_z(z), y \sim p_y(y)}[\log(1 - D(G(z|y)|y))].
\end{equation}

This conditioning enables CGANs to generate data that is not only realistic
 but also conforms to specified characteristics.
 In the context of regression tasks, \( y \) can represent numerical values, 
 prediction bounds, or parameters 
 that directly influence the generated data.
 For example, if the goal is to predict the mass of a particle, \( y \) could 
 include quantum features of the particle such as spin, charge, or isospin, 
 ensuring that the generated data remains consistent with these attributes.

The primary advantage of CGANs in particle physics is their ability  
to incorporate domain-specific constraints, ensuring that generated data adheres to 
the physical laws and characteristics of the problem. 
GANs and CGANs have become important tools in 
the field of particle physics, enabling researchers to simulate and analyze 
complex phenomena. 

Some notable applications include:
\begin{itemize}
    \item \textbf{Data Augmentation:} GANs  can be employed to generate synthetic datasets for rare events in HEP experiments, such as collisions in particle accelerators.
    \item \textbf{Detector Simulation:} GANs can assist in simulating particle trajectories in detectors, reducing computational costs compared to traditional Monte Carlo methods.
    \item \textbf{Quantum System Simulation:} GANs are capable of aiding in modeling quantum systems by generating data consistent with experimental results.
    \item \textbf{Hadronic Properties:} In this study, CGAN frameworks were employed to enhance the limited data on mesons and to predict their masses and decay widths by conditioning on quantum properties.
\end{itemize}

The section \ref{aug} provides a detailed explanation of data augmentation using CGANs.
 In this work, we trained the CGAN using the Adam optimizer for both the generator and discriminator with a learning rate of 0.002. 
The generator was trained with a custom regression loss function based on the log-cosh loss.
This choice helped the generator produce outputs that closely match the real meson properties,
improving regression performance while maintaining adversarial training. 
The discriminator was optimized using the binary cross-entropy loss
which encourages it to classify real versus generated samples correctly. 
Each model was trained for 1000 epochs with a batch size of 32.
To improve training stability, we applied dropout in both the generator and discriminator networks and used the Tanh activation function in all hidden layers.

It is worth mentioning that we applied the bagging technique to the CGAN model 
to aggregate predictions from multiple instances, 
thereby quantifying uncertainty in the generated outputs and 
improving the overall stability of the predictions. 
In this method, we set up $N$ independent CGANs: $\text{CGAN}_1, \text{CGAN}_2, \dots, \text{CGAN}_N$,
with $N$ being the number of models. 
Random sampling from the training data generates subsets, 
$\text{training data}_1, \text{training data}_2, \dots, \text{training data}_N$, 
which are used to train each CGAN model. 
This process, called bootstrapping, allows each model to learn different representations of the data.
After training, each CGAN produces predictions, 
denoted as $P_1, P_2, \dots, P_N$. 
The final output is obtained by averaging these predictions, 
$ \hat{P} = < \hat{P_{i}} > \equiv N^{-1}\Sigma_{i}\hat{P_{i}}$.
The variance across the predictions provides a measure of uncertainty, 
with higher variability indicating greater uncertainty in the model's output. 
This entire process, including bootstrapping and aggregation, is referred to as bagging \cite{Fujimoto:2021zas}.

\begin{figure}[h!]
	\begin{center} 
	\includegraphics[width=0.9\textwidth]{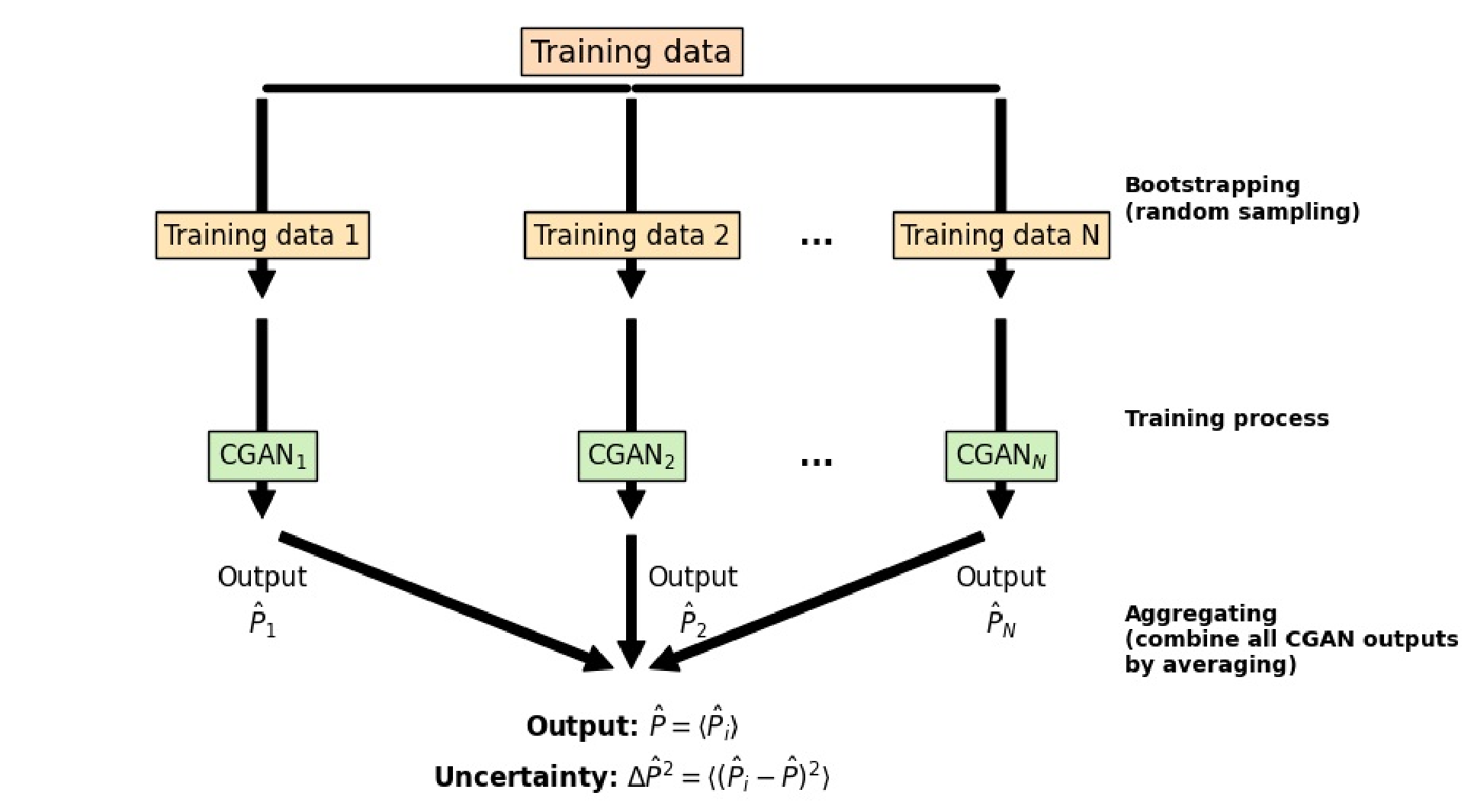}
	\end{center}
	\caption{Illustration of the bagging technique applied to the CGAN model. 
		The method involves training $N$ independent CGAN models on different subsets of the training data, 
		which are created through random sampling (bootstrapping).}
	\label{Plt2}
\end{figure}

In this work, we choose $N=10$ CGAN models.
The bagging technique is known to reduce the variance of the model predictions, 
which helps to reduce overfitting by averaging out fluctuations in the individual models' outputs. 
This leads to more stable and reliable predictions, 
especially in the presence of noisy data. 
The final output $ \hat{P} $ is less prone to extreme predictions compared to a single model.
The approach is  well illustrated in \textbf{Fig. \ref{Plt2}}. 

\section{Data augmentation and preprocessing}\label{aug}

As the NNs grow in complexity and scale, 
training leading-edge models requires vast amounts of data.
However, producing such data is often both resource-intensive and time-consuming. 
To manage this, one can either enhance the existing dataset with additional descriptive variables or mitigate data scarcity by artificially expanding the dataset through creation of new instances, bypassing the need for resource-intensive data generation. 

These approaches are collectively referred to as data augmentation techniques in ML applications. 
The first category of these methods, often called feature generation or feature engineering, 
is applied at the instance level. 
It involves creating new input features to provide more meaningful data for the algorithm, 
enhancing its ability to learn effectively. 

The second category of methods operates at the dataset level and 
can generally be divided into two main approaches. 
The first approach is known as real data augmentation, which involves making 
slight modifications to real data to create new samples. 
For instance, techniques like rotation or zooming are commonly used to augment image datasets. 
The second approach is synthetic data augmentation, where new data is generated entirely from scratch. 
This includes traditional sampling techniques and advanced generative models, 
such as GANs, which are capable of producing highly realistic synthetic datasets. 
Thus, by generating the synthetic data samples that retain key features or distributions of the original data, the synthetic data augmentation helps improve model generalization, enhances predictive performance, and facilitates the discovery of meaningful patterns in both experimental and simulated datasets.

In particle physics, the collection of experimental data is 
often both time-intensive, laborious and costly. 
Large-scale experiments, such as those conducted with particle colliders (e.g., the LHC), produce enormous volumes of data that require extensive preprocessing, analysis, 
and refinement to uncover meaningful insights. 

In LHC experiments, the data augmentation approach are introduced to accelerate simulation workflows. 
This method employs a generative deep learning model to transform collision events from an analysis-specific generator-level representation into their corresponding reconstruction-level representation \cite{Chen:2020uds}. 
Ref. \cite{Fujimoto:2021zas} augmented the training data using noise fluctuations corresponding to observational uncertainties. 
They suggest that the data augmentation could be an effective technique for reducing the possibility of overfitting without the need to adjust the NN architecture, such as by inserting dropout.

Applying deep learning methods to hadron physics may present several challenging problems. 
While PDG \cite{ParticleDataGroup:1986kuw,ParticleDataGroup:2024cfk} 
has cataloged hundreds of mesonic and baryonic states, 
the relatively limited number of known hadrons can pose significant challenges 
for advanced deep learning models. 
This scarcity of data may hinder the training process, 
potentially affecting the model's ability to generalize and make accurate predictions, 
especially when addressing complex phenomena in hadron physics.  

Two methods have been introduced for augmenting the hadronic data so far. 
The first involves adding and subtracting experimental mass errors from their central values while keeping the quantum numbers of hadrons fixed, 
effectively resampling the training data twice. The second method utilizes Gaussian noise resampling, generating random data points based on a Gaussian probability density function derived from the dataset's mean values and errors. This approach allows for up to nine replications of the hadronic data, 
with quantum numbers remaining constant throughout \cite{Bahtiyar:2022une, Bahtiyar:2022wph}.

Generative models, such as CGANs, present a powerful option for data augmentation in particle physics, particularly for generating synthetic hadronic data. 
A CGAN framework accomplishes this by producing synthetic samples 
conditioned on specific parameters or features, 
ensuring that the generated data adheres to the desired properties and 
aligns with the underlying physical characteristics. 

	In this study, we focused on mesons, both ordinary and exotic, whose quark compositions and quantum numbers, including isospin ($ I $), angular momentum ($ J $), parity ($ P $), $ G $-parity, and $ C $-parity, have been determined and confirmed by PDG \cite{ParticleDataGroup:1986kuw,ParticleDataGroup:2024cfk}. 

To handle potential ambiguities arising from mesons with identical quark structures and quantum numbers but differing masses, we introduce an additional feature referred to as the higher state ($ h $). It is important to emphasize that $ h $ is not a real quantum number but serves solely to differentiate particles with similar properties but varying masses. For instance, $\rho(770)$ and $\rho(1450)$ share identical input features but differ in mass. Thus, they are assigned $ h $ values of $ 0 $ and $ 1 $, respectively, allowing them to be recognized as distinct entities by machine learning algorithms. The range of $ h $ can vary from $ 0 $ to $ 10 $, depending on the number of similar mesons. 

The input vector for each meson is constructed as follows:

\begin{eqnarray}
	\label{eq}
	\vec{v} = (d,\bar{d},u,\bar{u},s,\bar{s},c,\bar{c},b,\bar{b},I,J,P,h,G,C),
\end{eqnarray}

where:
\begin{itemize}
    \item $ d, \bar{d}, u, \bar{u}, s, \bar{s}, c, \bar{c}, b, \bar{b} $ represent the presence (1) or absence (0) of each quark or antiquark in the meson's composition.
    \item $ I $ denotes the isospin, which can take values of $ 0 $, $ \frac{1}{2} $, or $ 1 $.
    \item $ J $ represents the total angular momentum, ranging from $ 0 $ to $ 6 $ in integer steps.
    \item $ P $ is the parity, which can be either $ -1 $ or $ +1 $.
    \item $ h $ is the higher state parameter, distinguishing mesons with identical quantum numbers but different masses.
    \item $ G $ and $ C $ represent $ G $-parity and $ C $-parity, respectively. If a meson is not an eigenstate of $ G $-parity or $ C $-parity, these values are set to $ 0 $.

\end{itemize}

Finally, the mass and decay width of the mesons are scaled using standard data preprocessing techniques to constrain the range of possible outputs. Since these parameters are dimensionful, they are divided by 1 MeV before scaling. This ensures that the input data is properly normalized for the ML algorithms.
The training dataset, comprising the mesons with accurately measured masses, 
and the test dataset, consisting of the mesons whose masses have yet to be determined. 
Furthermore, we extended this categorization considering the decay width of the mesons. 
In a similar manner, the mesons with all features including quark compositions, quantum numbers, 
$ h $, mass and decay width fully determined, were classified into the training dataset. 
Conversely, the mesons with unclear or undetermined decay 
width values were assigned to the test dataset.
This approach ensures a robust division of data, enabling focused training on well-characterized mesons while reserving those with incomplete information for testing and evaluation.
It is necessary to mention that, the mass and decay width of 
the mesons are expressed in units of MeV. 
However, number of the mesons in our training dataset is limited, 
necessitating expansion to support high-level deep learning architectures.
To manage this, we looked for an effective and professional approach 
to generate synthetic mesonic samples. 
The CGAN framework proved instrumental in achieving this goal, 
producing meaningful and reliable augmented data. 

For this purpose, CGAN takes the training data along 
with the associated experimental uncertainties in the mass or decay width. 
The augmentation Strategy can be explained as follow,

\begin{enumerate}

\item 
The mass and decay width values for mesons 
(e.g., the $\eta$ meson with mass $547.862 \pm 0.017 \,~\text{MeV}$ and 
width $1.31 \pm 0.05 \,~\text{MeV}$) are sampled within their experimental uncertainty ranges using a normal distribution. This approach allows the CGAN model to learn the variability within the experimental bounds.

\item 
Mesons lacking uncertainty data for mass or decay width are excluded from the augmentation process, as uncertainty sampling cannot be applied in such cases.

\item 
The CGAN is conditioned on fixed meson properties, such as quark content and quantum numbers ($I$, $J$, $P$, $C$, $G$), ensuring that the synthetic data maintains physical consistency while varying mass and decay width.

\item 
By sampling within uncertainty ranges, the CGAN learns the joint distribution of mass, decay width, and constant features, generating synthetic mesons that are consistent with both experimental data and uncertainties.
	
\end{enumerate}

\begin{figure}[h!]
	\begin{center} 
		\includegraphics[width=0.9\textwidth]{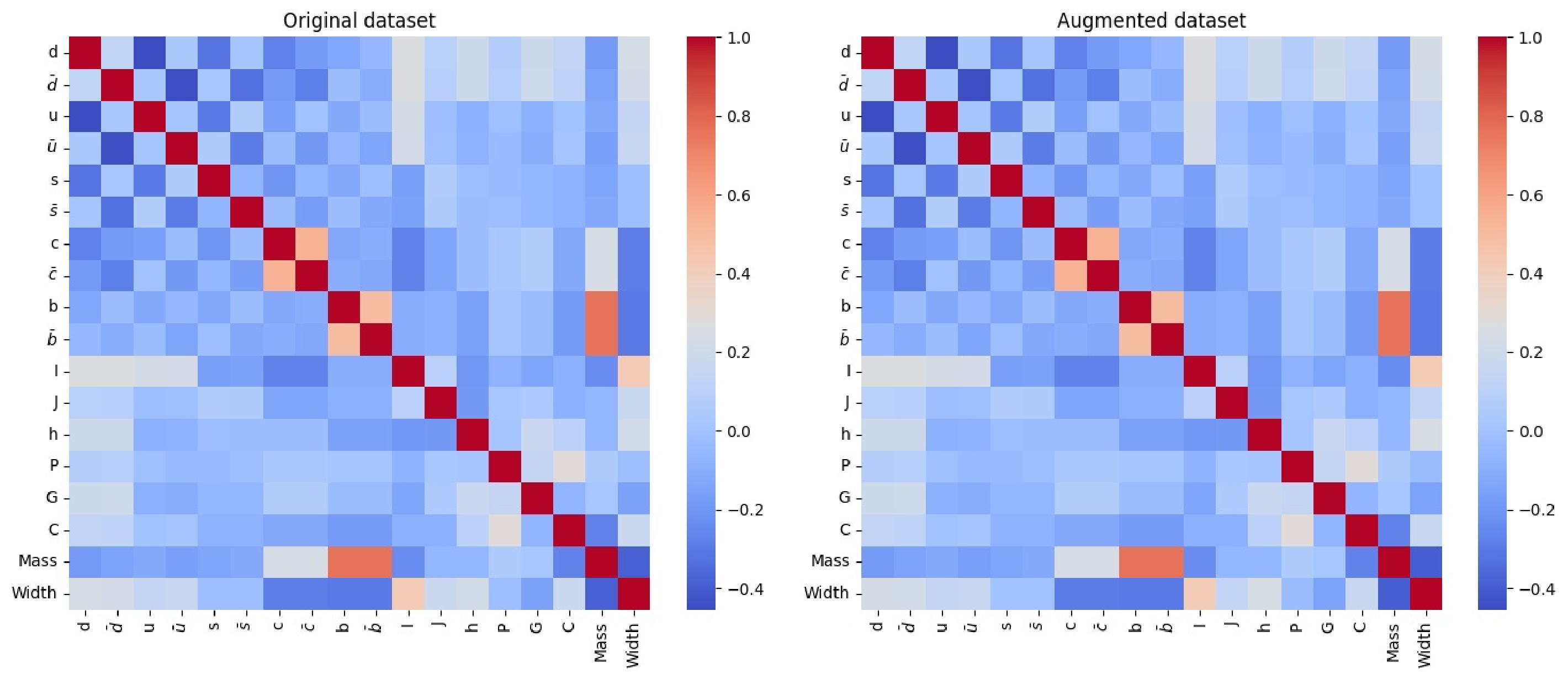}
	\end{center}
	\caption{Side-by-side comparison of heatmaps for the original and augmented meson datasets, 
		demonstrating how the augmented data preserves the patterns and distributions of the original dataset.}
	\label{Plt3}
\end{figure}

Consequently, we obtained synthetic training data that preserves the key properties of mesons, 
generated using one of the most effective generative models. 
The mesonic dataset was expanded by a factor of five, 
significantly increasing its size to enhance model training and analysis.
\textbf{Fig. \ref{Plt3}}, shows a side-by-side comparison of the heatmaps for the original and augmented meson datasets, 
providing strong evidence that the augmented data preserves the patterns and distributions of the original dataset. 
This comparison highlights the effectiveness of our CGAN framework in augmenting meson data 
and demonstrates the reliability of the augmentation method.

Following the data augmentation process, the next step involves scaling the mass and decay width values in the dataset. 
This is achieved using standard scaling techniques commonly employed in data science, 
such as normalization or standardization, 
to ensure that these values fall within a constrained range. 
Scaling not only aids in stabilizing the training process but also helps improve the model's ability to learn effectively from the data. 
Since the mass and width parameters are dimensionful, 
they are divided by 1 MeV to make them dimensionless before applying scaling. With this step completed, 
the dataset preparation process is finalized. 
At this stage, the input data is fully prepared and ready 
to be fed into our CGAN model to initiate the training process.

\section{Discussion}

In this part we give  and explain  our numerical results obtained using our CGAN model. 
We aim to predict the masses of several well-known, light mesons and exotic mesons, 
as well as the decay widths of mesons whose values have not yet been experimentally determined.
The numerical results for the predicted masses and decay widths, 
along with the corresponding mean errors, are presented in 
\textbf{Tables \ref{tab:wellmass}} to \textbf{\ref{tab:width2}}
The uncertainty in the predicted results has been determined using the bagging method described in Sec. \ref{GAN}.

\subsection{Ordinary or exotic mesons: an ongoing debate}

In contrast to ordinary mesons, tetraquarks are considered exotic mesons that fall outside the traditional framework of the quark model.
While ordinary mesons consist of a single quark-antiquark pair, 
tetraquarks contain four valence quarks, two quarks and two antiquarks, 
leading to their classification as more complex hadrons. 
Despite their unconventional structure, tetraquarks are still categorized as mesons 
because they consist of an equal number of quarks and antiquarks, adhering to the basic definition of mesons. 
Although the quark model was first introduced by Gell-Mann in 1964 \cite{Gell-Mann:1964ewy}, 
he also proposed the possibility of more exotic hadrons, beyond the conventional quark-antiquark structure of mesons and the three-quark structure of baryons. 
Since then, many mesons both ordinary and exotic have been discovered, 
yet the quark content of some remains ambiguous. 
This uncertainty raises the question of whether certain mesons should be classified as conventional ($q\bar{q}$) mesons or whether they may represent a new class of exotic hadrons, such as tetraquarks. 

In this study, we aim to predict the mass of several challenging mesons, including $a_0(980)$, $f_0(980)$, $D_{s0}^*(2317)^\pm$ and $D_{s1}(2460)^\pm$ using our CGAN model based on two key assumptions. 
first, we consider the $q\bar{q}$ structure, and then explore possibility of 
the $q\bar{q}q\bar{q}$ structure for their quark contents. 
 \textbf{Table \ref{tab:wellmass}} illustrates the numerical results compared to 
 the experimental \cite{ParticleDataGroup:2022pth},  
as well as  our previous  DNN results\cite{Malekhosseini:2024eot}. 
According to the \textbf{Table \ref{tab:wellmass}}, 
our CGAN  prediction for the mass of $a_0(980)$ 
in the $u\bar{u}$ configuration yields $983 \pm 44 \,~\text{MeV}$,
which is significantly closer to the experimental value of $980 \pm 20 \, \text{MeV}$ 
than our previous DNN result of $998 \pm 94 \,~\text{MeV}$. 
Such notable prdiction is also obtained when the $u\bar{s}\bar{u}s$ configuration is supposed for $a_0(980)$. 
In fact, the masses estimated by the CGAN 
are not only closer to the experimental values but also exhibit 
smaller uncertainty ranges compared to the DNN results.   
If we examine the other particles in \textbf{Table \ref{tab:wellmass}}, 
we find that our CGAN predictions outperform the DNN results for both ordinary and exotic assumptions.
Besides our CGAN model results suggest that, based on the predicted mass distributions, 
the possibility of these mesons being tetraquarks cannot be ruled out.
\footnote{For the machine learning model, pure quark–antiquark configurations 
such as $u\bar{u}$ or $d\bar{d}$ were used as simplified inputs. 
However, the physical isospin eigenstates are given by 
$(u\bar{u} - d\bar{d})/\sqrt{2}$ for $I=1$ and 
$(u\bar{u} + d\bar{d})/\sqrt{2}$ for $I=0$.}
Moreover, our CGAN predictions are generally consistent with QCD-based 
theoretical analyses. For instance, QCD sum-rule studies of $a_0(980)$ 
and $f_0(980)$ \cite{Wang:2005cn} 
 indicate that $a_0(980)$ 
 meson has an energy of
1115~MeV,
and finite-volume lattice QCD 
calculations \cite{Alexandrou:2017itd} present 
 a candidate for the
masses near 950-1020~MeV, in agreement with our results within uncertainties. 
Similarly, the $D_{s0}^*(2317)$ and $D_{s1}(2460)$ meson masses predicted by 
our model are compatible with HQET sum-rule  values of 
2420 and 2610 MeV, 
\cite{Dai:2004yk}
and lattice QCD calculations \cite{Bali:2017pdv}, which yield masses 
around 2348 and 2451 ~MeV. These comparisons suggest that the CGAN approach captures 
the main mass features predicted by established QCD-like methods while providing smaller uncertainties.

\subsection{Light mesons}
The mass of $f_0(500)$ is known to lie within the range of $400$ to $800$ MeV (see \textbf{Table \ref{tab:lightmass}}).
Our CGAN prediction yields a mass of $547 \pm 102 \,~\text{MeV}$,
which falls well within this range. In comparison, the DNN model estimates the mass to be
$759 \pm 134 \,~\text{MeV}$, which is closer to the upper end of the known range.
While both predictions are within the experimentally expected range, 
the CGAN prediction provides a value closer to the central region of the mass range. 
The smaller uncertainty range in the CGAN prediction also suggests a more confident 
estimation compared to the DNN result.
Our CGAN predictions for the masses of $K_{4}(2500)_0,\bar{K_{4}}(2500)_0$ and $K_{4}^{\pm}(2500)$
are in good agreement with the experimental values. 
Our CGAN predictions for the mass of  $K_{2}(1580)_0,\bar{K_{2}}(1580)_0$ is $1702\pm {93}$ 
and for the mass of $K_{2}^{\pm}(1580)$ is $1733\pm {98}~\text{MeV}$, 
which are somewhat higher than the experimental value. 
Despite this discrepancy, the CGAN predictions provide valuable insight into the mass range, 
though further refinement may be needed for more accurate alignment with the experimental mass.
Additionally, our CGAN prediction for the mass of $f_0(500)$ is 
in good agreement with various QCD-based theoretical studies. 
QCD sum-rule analyses that assume tetraquark structure or mixing 
\cite{Cid-Mora:2023hwd,Agaev:2017cfz} yield masses in the range 
of 480-770~MeV, while QCD sum rules in Ref.~\cite{Wang:2015uha} give $660 \pm 6$~MeV. 
Lattice QCD simulations employing 
two-pion interpolating operators \cite{Howarth:2015caa} 
report $609 \pm 80$~MeV. These results collectively 
indicate that the $f_0(500)$ mass lies well below 800~MeV, 
consistent with our CGAN estimate of $547 \pm 102$~MeV and 
supporting its interpretation as a broad scalar resonance 
with a complex internal structure.

\subsection{Exotic mesons}
In this section, we present our CGAN predictions for the mass of several exotic states
comparing them to experimental measurements as well as the results obtained from our previous DNN model (see  \textbf{Table \ref{tab:exomass}}).
The CGAN model predicts a mass of $3716\pm {140}~\text{MeV}$ for $\chi_{c1}(3872)$.
While this prediction is not an exact match to the experimental value of $3871.65\pm {0.06}~\text{MeV}$,
it is still much closer than the DNN prediction of $2944 \pm{177}~\text{MeV}$, 
highlighting the superior accuracy of the CGAN model in this instance.
For states such as $\psi(4230)$, $\psi(4360)$ and $\psi(4660)$, 
our CGAN model yields predictions that are notably closer to 
the experimental values than the DNN predictions.  
Also, the CGAN strongly predicts the mass of $Z_c(3900)^\pm$ state, 
where the experimental mass is reported as $3887.1\pm {2.6}~\text{MeV}$, 
The CGAN prediction of  $3876 \pm{210}~\text{MeV}$, 
is very close to this value, demonstrating its ability to accurately capture the mass of this state. 
While there is some uncertainty in the CGAN prediction, it is still within a reasonable range of the experimental measurement.
This trend persists for other exotic states listed in the table.
For instance, the CGAN estimates the mass of $Z_b(10650)^\pm$ to be $10675\pm{691}~\text{MeV}$, 
which aligns closely with the experimental value of $10652.2\pm{1.5}~\text{MeV}$.

These comparisons highlight the advanced performance of the CGAN model, 
especially when combined with the augmentation technique, compared to our previous DNN model. 
The augmentation of the training data plays a critical role in enhancing the model’s ability 
to learn complex patterns and generalize more effectively. 
By artificially expanding the dataset, we provide the model with a more diverse set of examples. 
This enriched data allows the CGAN model to better capture the underlying relationships between features, resulting in more accurate and robust predictions.
Furthermore, the augmentation technique also contributes to reducing overfitting, 
which is often a challenge in deep learning models with limited data. 
By exposing the CGAN to a broader range of training examples, 
we improve its ability to handle unseen data with greater precision, 
particularly for the challenging task of predicting the masses and width of exotic states.
Thus, the CGAN model, with the added benefit of data augmentation, 
outperforms the DNN not only in terms of prediction accuracy 
but also in its ability to generalize and handle complex patterns with higher precision.
Furthermore, our CGAN predictions for exotic 
states are generally consistent with theoretical studies. 
For example, lattice calculations for $X(3872)$ indicate a 
mass of $3890 \pm 30$~MeV \cite{Chiu:2006hd}.
Moreover, QCD sum-rule for $\psi(4230)$, $\psi(4360)$, 
and $\psi(4660)$ predict masses around 4217, 4263, and 4359~MeV \cite{Bhavsar:2020pog}, 
respectively, consistent with experimental values within 
uncertainties. QCD sum-rule studies also support tetraquark 
or molecular interpretations for  $Z_b(10610)\sim10597-10609$ MeV, 
and  $Z_b(10650)\sim10638-10648$  MeV \cite{Yang:2011rp}, with predicted masses close to the measured ones. 
Overall, these comparisons indicate that our CGAN model captures the main mass 
features of exotic states and performs notably better than the DNN, especially 
when combined with data augmentation.

 \begin{table}[h!]
\begin{center}

\renewcommand{\arraystretch}{1.4}
\scalebox{0.66}{
\begin{tabular}{|c|c|c|c|c|c|}\cline{1-6}
 Meson&  $I^G\,(J^{PC})$	 &   Exp. Mass (MeV)  \cite{ParticleDataGroup:2022pth}  &  quark content & DNN\cite{Malekhosseini:2024eot} & CGAN model\\ \hline
 \hline
\multirow{2}{*}{$a_0(980)$}	 &\multirow{2}{*}{ $1^-\,(0^{++})$}	& \multirow{2}{*}{$980\pm 20$} & $u\bar{u}$       &  $998\pm {94}$&$983\pm {44}$\\ \cline{4-6}
& & & $u\bar{s}\bar{u}s$ $(K\overline{K})$   & $1069\pm {224}$& $994\pm {142}$\\ \hline
\multirow{2}{*}{$f_0(980)$}	 &\multirow{2}{*}{ $0^+\,(0^{++})$}	& \multirow{2}{*}{$990\pm 20$} & $d\bar{d}$   &  $883\pm {45}$ & $805\pm {60}$\\ \cline{4-6}
& & & $d\bar{s}\bar{d}s$ $(K\overline{K})$   & $1086\pm {68}$&$1032\pm {27}$\\ \hline
\hline
\multirow{2}{*}{$D_{s0}^*(2317)^\pm$}	 &\multirow{2}{*}{ $0\,(0^+)$}	& \multirow{2}{*}{$2317.8\pm0.5$} &$\pm {}$ $c\bar{s}$   & $2343\pm {169}$&$2312\pm {49}$ \\ \cline{4-5}\cline{6-6}
& & & $c\bar{u}u\bar{s}$ $(DK)$ & $2511\pm {334}$&$2344\pm {70}$\\ \hline
\multirow{2}{*}{$D_{s1}(2460)^\pm$}	 &\multirow{2}{*}{ $0\,(1^+)$}	& \multirow{2}{*}{$2459.5\pm 0.6$} & $c\bar{s}$    &$2442\pm {218}$&$2447\pm {48}$\\ \cline{4-5}\cline{6-6}
& & & $c\bar{u}u\bar{s}$ $(D^*K)$  &$2748\pm {504}$&$2365\pm {132}$\\ \hline
\end{tabular}
}
	\caption{Predictions of our CGAN model for the masses of four well-known conventional mesons compared to their corresponding tetraquark structures (in MeV). Results are presented alongside experimental data \cite{ParticleDataGroup:2022pth} and our previous DNN model predictions in Ref. \cite{Malekhosseini:2024eot}.}\label{tab:wellmass}
 \end{center}
\end{table}

 \begin{table}[h!]
\begin{center}

\renewcommand{\arraystretch}{1.4}
\scalebox{0.68}{
\begin{tabular}{|c|c|c|c|c|c|}\cline{1-6}
 Meson&  $I^G\,(J^{PC})$	 &  Exp. Mass (MeV)  \cite{ParticleDataGroup:2022pth}  &  quark content  &  DNN\cite{Malekhosseini:2024eot}& CGAN model  \\ \hline
\hline
$f_0(500)$ & $0^+\,(0^{++})$ & $400-800$  & $d\bar{d}$    &$759\pm {134}$ &$547\pm {102}$\\ \hline

$K_{4}(2500)_0,\bar{K_{4}}(2500)_0$ & $1/2\,( 4^-)$ & $2490\pm {20}$  & $d\bar{s}, s\bar{d}$   &$2308\pm {35}$&$2397\pm {55}$ \\ \hline

$K_{4}^{\pm}(2500)$ & $1/2\,( 4^-)$ & $2490\pm {20}$  & $u\bar{s}, s\bar{u}$  &$2298\pm {25}$&$2404\pm {89}$ \\ \hline

$K_{2}(1580)_0,\bar{K_{2}}(1580)_0$ & $1/2\,( 2^-)$ & $1580$  & $d\bar{s}, s\bar{d}$   &$1646\pm {20}$&$1702\pm {93}$ \\ \hline

$K_{2}^{\pm}(1580)$ & $1/2\,( 2^-)$ & $1580$  & $u\bar{s}$  &$1653\pm {23}$&$1733\pm {98}$ \\ \hline

\end{tabular}
}	\caption{Predictions of our CGAN model  for the mass of some light mesons (in units of MeV).
Results are presented alongside experimental data \cite{ParticleDataGroup:2022pth} and our previous DNN model predictions in Ref. \cite{Malekhosseini:2024eot}.}\label{tab:lightmass}
 \end{center}
\end{table}

 \begin{table}[h!]
\begin{center}
\renewcommand{\arraystretch}{1.4}
\scalebox{0.59}{
\begin{tabular}{|c|c|c|c|c|c|}\cline{1-6}
 Meson&  $I^G\,(J^{PC})$	 &  Exp. Mass (MeV) \cite{ParticleDataGroup:2022pth}  & quark content & DNN\cite{Malekhosseini:2024eot}& CGAN model   \\ \hline
 \hline
$\chi_{c1}(3872)$	 
& $0^+\,(1^{++})$	& $3871.65\pm 0.06$  & $c\bar{u}\bar{c}u$ $(D^0\,\bar{D}^{*0})$ & $2944\pm {177}$ & $3716\pm {140}$ \\ \hline
$\psi(4230)$ & $ 0^-\,( 1^{--})$ & $4222.5\pm 2.4$  & $c\bar{s}\bar{c}s$ $(D_s\,\bar{D}_s)$ &$3303\pm {175}$& $4076\pm {190}$ \\ \hline
$\psi(4360)$ & $ 0^-\,( 1^{--})$ &  $4374\pm 7$  &  $c\bar{u}\bar{c}u$ $(D_1\,\bar{D}^*)$ &  $3190\pm {184}$& $4204\pm {175}$ \\ \hline
$\psi(4660)$ & $0^-\,( 1^{--})$ & $4630\pm 6$  &  $c\bar{u}\bar{c}u$ $(f_0(980)\,\psi^\prime)$ & $3300\pm {79}$& $4456\pm {251}$ \\ \hline
$Z_c(3900)^\pm$ & $ 1^+\,( 1^{+-})$	&$3887.1\pm 2.6$  & $\bar{c}uc\bar{d}$ $(D\,\bar{D}^*)$ &  $3676\pm {183}$ & $3876\pm {210}$\\ \hline 
$Z_c(4200)^\pm$ & $1^+\,( 1^{+-})$	&  $4196^{+35}_{-32}$  & $\bar{c}uc\bar{d}$ & $3981\pm {195}$& $4172\pm {103}$\\   \hline

$Z_c(4430)^\pm$ & $1^+\,( 1^{+-})$	& $4478^{+15}_{-18}$ & $\bar{c}uc\bar{d}$  $(D_1 D^*\,,\,D_1^\prime D^*)$    &$4052\pm {197}$& $4011\pm {676}$\\  \hline

$Z_b(10610)^\pm$ & $ 1^+\,( 1^{+-})$	& $10607.2\pm2.0$  & $b\bar{d}\bar{b}u$  $(B \bar{B}^*)$ & $8918\pm {447}$& $10486\pm {685}$ \\  \hline
$Z_b(10650)^\pm$ & $1^+\,( 1^{+-})$	& $10652.2\pm1.5$  &$b\bar{d}\bar{b}u$ $(B^* \bar{B}^*)$ &   $9103\pm {159}$ & $10675\pm {691}$ \\ \hline

$Z_{cs}(4220)^+$ & $1/2\,(1^+)$ & $4216^{+50}_{-40}$  & $u\bar{c}\bar{s}c$ &$3054\pm {182}$ & $4144\pm {138}$\\ \hline
$R_{c0}(4240)$ & $1^+\,( 0^{--})$ & $4239^{+50}_{-21}$  & $c\bar{u}\bar{c}u$ & $3760\pm {469}$& $4096\pm {102}$ \\ \hline

\end{tabular}}
	\caption{Predictions of our CGAN model for the mass of some exotic mesons (in units of MeV). 
Results are presented alongside experimental data \cite{ParticleDataGroup:2022pth} and our previous DNN model predictions in Ref. \cite{Malekhosseini:2024eot}.}\label{tab:exomass}
\end{center}
\end{table}

\subsection{Fully heavy tetraquarks}
The experiments at the LHC, conducted by collaborations such as LHCb, ATLAS, 
and CMS, have played a transformative role in advancing our understanding of exotic hadronic states. 
Recently, resonances such as \(X(6200)\), \(X(6600)\), \(X(6900)\), and \(X(7300)\) 
have been observed in the invariant mass spectra of di-\(J/\psi\) and \(J/\psi \psi^\prime\). 
These discoveries have enhanced both theoretical and experimental efforts 
to explore fully-heavy tetraquark systems composed of \(c\)- and \(b\)-quarks \cite{LHCb:2020bwg, Bouhova-Thacker:2022vnt, CMS:2023owd}.

Fully-heavy tetraquarks, characterized by their unique structure consisting only of heavy quarks and antiquarks (e.g., \(cc\bar{c}\bar{c}\) or \(bb\bar{b}\bar{b}\)), represent a distinct class of multiquark states. 
Unlike conventional mesons and baryons, these exotic states open up a new perspective on the dynamic of QCD  
in the non-perturbative regime. Their existence was theorized decades ago, 
tracing back to the pioneering work of Gell-Mann and Zweig during the development of the quark model
\cite{Iwasaki:1975pv, Chao:1980dv}. 
Theoretical predictions for these states first emerged from non-relativistic potential models, 
QCD sum rules, and lattice QCD calculations \cite{Ader:1981db, Lipkin:1986dw, Zouzou:1986qh}.

The LHCb collaboration's landmark observation of structures in the di-\(J/\psi\) 
spectrum provided the first compelling evidence for fully-charmed tetraquark candidates. 
These resonances were identified with masses that significantly exceed the thresholds of conventional charmonium states, 
These findings confirmed their exotic nature, 
with subsequent analyses by the ATLAS and CMS collaborations providing high-statistics data that helped 
constrain their properties, including masses and decay widths. \cite{CMS:2023owd}.

These experimental achievements have paved the way for future searches, 
including investigations into fully-bottom (\(bb\bar{b}\bar{b}\)) and mixed heavy-quark configurations (e.g., \(bc\bar{b}\bar{c}\)). 
The upcoming high-luminosity LHC (HL-LHC) and Belle II experiments are expected to increase sensitivity to these states, 
potentially revealing new tetraquark families.

The theoretical understanding of fully-heavy tetraquarks has evolved significantly \cite{Agaev:2023gaq,Agaev:2023ruu,Agaev:2023rpj,Agaev:2023ara,Agaev:2023tzi,Agaev:2024pej,Agaev:2024pil,Agaev:2024xdc,Agaev:2024wvp,Agaev:2024mng,Agaev:2024qbh,Agaev:2024uza}. 
Early studies used diquark-antidiquark configurations to estimate masses and binding energies of these states. 
Using QCD sum rules, studies have systematically explored quantum numbers such as \(J^{PC} = 0^{++}, 1^{++},\) and \(2^{++}\). 
These analyses consistently predict masses above the dissociation thresholds of two quarkonium states, 
indicating their dominant decay modes \cite{Berezhnoy:2011xn, Chen:2016jxd, Wang:2017jtz, Anwar:2017toa}.

Despite significant progress, several open questions remain. 
For instance, the stability of fully-heavy tetraquarks against strong decays has yet to be comprehensively understood. 
Additionally, the role of chromoelectric and chromomagnetic interactions in binding these systems is still debated. 
Recent theoretical studies have provided estimates for the masses and decay properties of fully-heavy tetraquarks. In contrast, we employed the CGAN framework to predict the masses of some of these tetraquarks, further enhancing our understanding of their properties.
\textbf{Table \ref{tab:exo_heavy}} presents CGAN-based predictions for fully-heavy tetraquark masses.
It should be noted that, $ T_{cc\bar{c}\bar{c}}$, 
the only heavy tetraquark in \textbf{Table \ref{tab:exo_heavy}} with an experimentally measured mass, 
serves as a benchmark for comparison. The experimentally measured mass is reported as $6899 \pm 12 ~\text{MeV}$.
Our CGAN estimate for this heavy tetraqurk is $6763 \pm{681}~\text{MeV}$. 
While it seems a difference between the experimental and predicted values, 
this prediction was particularly challenging due to the complex nature of fully heavy tetraquarks.
Nonetheless, the predicted result demonstrates the potential of our CGAN framework in making non-trivial mass prediction,
and further refinements could bring the estimate closer to experimental value. 
The remaining CGAN predictions for heavy tetraquarks in \textbf{Table \ref{tab:exo_heavy}}
are presented alongside theoretical estimates for comparison.

Besides the QCD sum-rule results listed in Table~\ref{tab:exo_heavy}, 
we also compared our predictions with a variety of other theoretical 
approaches. For example,for the fully-bottom tetraquark $T_{bb\bar{b}\bar{b}}$ with $J^{PC}=0^{++}$, 
our CGAN prediction ($18388\pm912$~MeV) lies within the broad range of QCD-inspired approaches. 
For example, QCD sum-rule calculations give $18130-18840$~MeV 
\cite{Wang:2021mma,Wang:2017jtz,Wang:2018poa,Agaev:2023wua}, 
while nonrelativistic constituent-quark models predict masses around 
$19200-19350$~MeV 
\cite{An:2022qpt,Wang:2019rdo,Liu:2019zuc,Zhang:2022qtp}, 
Bethe-Salpeter equations yield $19205$~MeV \cite{Li:2021ygk}, 
relativistic quark models give $19201-19255$~MeV \cite{Faustov:2020qfm,Lu:2020cns}, 
 Monte-Carlo or diquark-based models predict 
$19199$~MeV \cite{Gordillo:2020sgc}, 
effective potential models 19154-19226~MeV \cite{Zhao:2020nwy}, 
and flux-tube models about 19329~MeV \cite{Deng:2020iqw}.

A similar level of agreement is observed for the  state 
$T_{bb\bar{c}\bar{c}}$. For the $J^{PC}=0^{+}$ 
state we obtain $13054\pm917$~MeV, compared with $12715-13383$~MeV 
from QCD sum rules \cite{Agaev:2024xdc,Agaev:2023tzi} and 
$12847-13039$~MeV from various nonrelativistic constituent-quark 
models \cite{An:2022qpt,Wang:2019rdo,Liu:2019zuc,Zhang:2022qtp}. 
Relativistic, Monte Carlo, diquark, flux-tube and chromomagnetic models 
also predict masses in the $12380-13630$~MeV interval 
\cite{Faustov:2020qfm,Gordillo:2020sgc,Mutuk:2022nkw,Deng:2020iqw,
Wu:2016vtq,Weng:2020jao,Chen:2020lgj}, consistent with our CGAN values.

For the $T_{bc\bar{b} \bar{c}}$ state with $J^{PC}=0^{++}$ our CGAN predicts 
$11880\pm592$~MeV this is close to the QCD sum-rule estimate $12697$~MeV 
\cite{Agaev:2024wvp} and to the lattice-QCD results 
$12820-13449$~MeV \cite{Yang:2021hrb}. 
Nonrelativistic constituent-quark models give $12760-12989$~MeV 
\cite{An:2022qpt}, $12854-12931$~MeV \cite{Liu:2019zuc}, and 
$12783-12966$~MeV \cite{Zhang:2022qtp}, while the relativistic 
quark model predicts $12813-12824$~MeV \cite{Faustov:2020qfm}. 
Monte Carlo methods yield $12534$~MeV \cite{Gordillo:2020sgc}.
Our CGAN estimate therefore sits between QCD sum rules and various effective 
potential or flux-tube models, 
demonstrating that the network captures the overall mass scale predicted by 
QCD-like effective approaches.

\begin{table}[h!]
	\begin{center}
		\renewcommand{\arraystretch}{1.4}
		\scalebox{0.59}{
			\begin{tabular}{|c|c|c|c|}\cline{1-4}
				Meson&  $I^G\,(J^{PC})$	 & Mass (MeV)   &CGAN model   \\ \hline
				\hline
				$ T_{cc\bar{c}\bar{c}}$	 
				& $0^+\,(0^{++})$	& $6899 \pm 12 $   \cite{ParticleDataGroup:2022pth}(Exp. Mass)  & $6763 \pm{681}$ \\ \hline
				$ T_{cc\bar{c}\bar{c}}^*$	 
				& $0^+\,(0^{++})$	& $7235 \pm 75 $ \cite{Agaev:2023rpj}  & $7014 \pm {371}$ \\ \hline
				
				$ T_{bb\bar{b}\bar{b}}$	 
				& $0^+\,(0^{++})$	& $18858\pm {50}$ \cite{Agaev:2023gaq} & $18388\pm {912}$ \\ \hline

				$T_{bb\bar{b}\bar{b}}^* $	 
				& $0^+\,(0^{++})$	& $18540\pm {50}$ \cite{Agaev:2023ara}& $19100\pm {965}$ \\ \hline

				$ T_{cc\bar{c}\bar{b}}$	 
				& $0^+\,(0^{+})$	& $9680\pm 102 $\cite{Agaev:2024uza}   & $9736\pm {681}$ \\ \hline

				$ T_{bb\bar{b}\bar{c}}$	 
				& $0^+\,(0^{+})$	 & $15697\pm {95}$\cite{Agaev:2024uza} & $15006\pm {857}$ \\ \hline

				$ T_{bb\bar{c}\bar{c}}$	 
				& $0^+\,(0^{+})$	& $12715\pm {86}$ \cite{Agaev:2023tzi} & $13054\pm {917}$ \\ \hline
				
				$T_{bb\bar{c}\bar{c}} $	 
				& $0^+\,(0^{-})$	& $13092\pm {950}$ \cite{Agaev:2024xdc}& $12774\pm {821}$ \\ \hline

				$ T_{bb\bar{c}\bar{c}}$	 
				& $0^+\,(1^{-})$	& $13092\pm {95}$ \cite{Agaev:2024xdc}& $11312\pm {980}$ \\ \hline

				$ T_{bb\bar{c}\bar{c}}$	 
				& $0^+\,(2^{+})$	& $12795\pm {950}$ \cite{Agaev:2024pil} & $11908\pm {726}$ \\ \hline

				$ T_{bc\bar{b}\bar{c}}$	 
				& $0^+\,(0^{++})$	& $12697\pm {90}$ \cite{Agaev:2024wvp} & $11880\pm {592}$ \\ \hline

				$ T_{bc\bar{b}\bar{c}}$		 
				& $0^+\,(0^{-+})$	& $----$ & $11806\pm {680}$ \\ \hline
				
				$ T_{bc\bar{b}\bar{c}}$	 
				& $0^+\,(1^{-+})$	& $----$ & $11699\pm {506}$ \\ \hline

				$ T_{bc\bar{b}\bar{c}}$		 
				& $0^+\,(1^{++})$	& $12715\pm {90}$ \cite{Agaev:2024mng} & $11783\pm {940}$ \\ \hline

				$ T_{bc\bar{b}\bar{c}}$		 
				& $0^+\,(2^{++})$	& $12700\pm {90}$ \cite{Agaev:2024qbh} & $11470\pm {303}$ \\ \hline
		\end{tabular}}
		\caption{Predictions of our CGAN model  for the mass of fully heavy tetraquark (in units of MeV). 
			Results are presented alongside theoretical prediction.}\label{tab:exo_heavy}
	\end{center}
\end{table}

\subsection{Meson width}
The widths of some exotic mesons remain poorly constrained, due to challenges in experimental measurements and the complex nature of these particles. 
As the understanding of exotic hadrons evolves, accurate predictions of their properties, 
including their decay widths, become essential in advancing our knowledge of strong interactions. 

In this section, we predict the decay widths of several exotic mesons, using our CGAN model. 
The CGAN approach, combined with augmented training data, 
provides an alternative to traditional methods, 
such as the DNN, offering the potential for improved predictions by 
learning intricate patterns in the data.

Although most theoretical studies on meson widths have concentrated on partial
decay channels rather than on the total width, there exist a few notable
calculations of the full width that allow for a direct comparison with our
predictions. For example,
Ref.~\cite{Agaev:2018fvz}, using the QCD two-point sum rule approach, estimates
the total width of the $a_{0}(980)$ meson as $(62.0\pm14.4)$~MeV, while the
values for the $f_{0}(980)$ meson in Ref.~\cite{Escribano:2002iv} lie in the
range $39$–$52$~MeV. For heavy states, Ref.~\cite{Jia:2024imm} calculates the
full width of the $D^{*}(2007)^{0}$ meson to be nearly $54$~keV.
In Table~\ref{tab:width1} we compare our CGAN predictions with these full-width
results and with experimental data, showing that our model yields widths of the
correct order of magnitude and, for most states, within or close to the quoted
ranges. This demonstrates the capability of the CGAN framework to go beyond
channel-specific calculations and provide non-trivial predictions for total
widths.

The experimental width of $a_0(980)$ meson, reported as $97 \pm 1.9 \pm 5.7~\text{MeV}$,
provides a reference point for comparing theoretical predictions.
Our previous DNN prediction was $113 \pm 28~\text{MeV}$, while the GAN
model predicts a width of $101 \pm 33~\text{MeV}$.
Even though the CGAN prediction is closer to the experimental value, 
both the DNN and CGAN results fall within an acceptable range considering the uncertainties.

When the $a_0(980)$ is assumed to be an exotic meson,
the CGAN model predicts a larger width of $ 210 \pm 80~\text{MeV}$. 
This increase in the predicted width when the $a_0(980)$ is modelled as an exotic meson, 
suggests that the assumption of its exotic nature has a notable impact on the predicted decay width. 
Exotic mesons are typically associated with more complex internal structures, 
which could lead to broader decay widths due to different decay channels or more complex dynamics.

The situation for $f_0(980)$ is somewhat different. 
The experimental width is estimated to lie within the range of $10$ to $100~\text{MeV}$. 
The DNN model obtained widths of $105 \pm 34~\text{MeV}$ and $120 \pm 58~\text{MeV}$,
when considering it as an ordinary meson and an exotic meson, respectively.
These predictions are slightly above the experimental range but 
still within a reasonable range when considering the uncertainty.
 Our CGAN framework predicts a width of $82 \pm 32~\text{MeV}$ for
the $d\bar{d}$ configuration, which falls comfortably within the experimental range.
for the $d\bar{s}\bar{d}s$ configuration (exotic meson),
the CGAN model predicts a width of $105 \pm 80~\text{MeV}$,
which is still within the expected experimental range, though towards the upper limit. 

While, Both of the CGAN prdictions for the ordinary and exotic states of $f_0(980)$,
are in close conformity with the experimental expectations,
The width for the exotic configuration is larger than the ordinary configuration, 
indicating that treating the meson as an exotic state leads to a broader predicted decay width, 
which is consistent with the expectation that exotic mesons may have more decay channels 
or more complex dynamics.

For $f_0(1370)$, the experimental width is estimated to fall between $200$ and $ 500 \text{MeV}$.
The DNN model obtained a width of $107 \pm 40~\text{MeV}$, 
and in comparison, our CGAN prediction gives $197 \pm 62~\text{MeV}$.
Despite the fact that both predictions are below the lower limit of the experimental range, 
the CGAN result is closer to the lower end of the expected width range.
This points to better agreement with the experimental data compared to our previous DNN prediction.

The predictions for other remaining mesons, presented in \textbf{Table \ref{tab:width1}},
show remarkable results when compared to experimental data.
For the $ D^*(2007)^0$, the experimental width is constrained to be less than $ 2.1~\text{MeV}$ 
at a $90\%$ confidence level, while the CGAN estimate is much lower at $ 1.6 \pm 0.8~\text{MeV}$, 
revealing a strong agreement with the experimental upper limit. 
Likewise, for the $ D_{s_0}^{*}(2317)^{\pm}$,
both the DNN and CGAN models predict widths around $ 3~\text{MeV}$, consistent with 
the experimental upper bound of $ 3.8~\text{MeV}$.
A smaller decay width is typically expected for conventional mesons, as seen in these models.
When an exotic interpretation of the $ D_{s_0}^{*}(2317)^{\pm}$, is considered,
it could involve more complex internal structure or interactions.
Such structure would naturally cause a broader decay width compared to conventional mesons.

The predicted width of this exotic state was obtained $ 47 \pm 23~\text{MeV}$ by the DNN.
The CGAN model predicts a decay width of $11.7 \pm 7 \text{MeV}$, 
which is still above the experimental upper bound but smaller than the prediction from the DNN model.
The larger predicted widths for the exotic interpretation of the $ D_{s_0}^{*}(2317)^{\pm}$ meson, 
especially the DNN model's prediction of $47 \pm 23~\text{MeV}$, 
suggest that this scenario does not align well with the experimental data, 
which supports a much smaller decay width. 

The experimental upper bound for the decay width of 
the $ D_{s1}(2460)^{\pm}$ meson is estimated to be less than $ 3.5~\text{MeV}$.
The CGAN and DNN predictions for the ordinary state of this meson show good consistency with the experiment.
However, the exotic prediction for the $ D_{s1}(2460)^{\pm} $, gives a width of $ 31 \pm 21~\text{MeV}$,
according to the DNN and $ 8.2 \pm 3~\text{MeV}$ according to the CGAN model.
Kindly note that the CGAN model shows an improvement compared to the DNN model, as its predicted decay width is smaller, 
but it is still larger than the experimental upper bound.
It can be implied that the tetraquark hypothesis for the $ D_{s1}(2460)^{\pm}$ meson may not be fully consistent with the experimental data,
similar to the $ D_{s_0}^{*}(2317)^{\pm}$ exotic prediction.
For the $\psi_2(3823)$, the CGAN prediction of $ 2.7 \pm 1.5~\text{MeV}$,
fits well with the experimental upper limit of $ 2.9~\text{MeV}$.

The experimental decay width of the $\eta_b(2s)$ state is estimated to be less than $24~\text{MeV}$.
The DNN model predicted the width of $ 54 \pm 24~\text{MeV}$ whearas, 
The CGAN prediction is obtained $ 12.8 \pm 4~\text{MeV}$.
The CGAN result is notably smaller and more precise than the DNN estimate. 
Moreover, it is closer to the expected experimental value. 

Finally, the predictions for the kaon resonances 
$ K_0^*(700)$ and $ K_0^{*}(1430) $ closely match the experimental measurements. 
For $ K_0^*(700)$, the CGAN estimation of $ 461 \pm 146~\text{MeV}$,
Corresponds closely to the experimental value of $ 468 \pm 30~\text{MeV}$.
Analogously, the CGAN result for $ K_0^{*}(1430) $,
is in good agreement with the experimental value of $ 270 \pm 80~\text{MeV}$,
with a prediction of $ 274 \pm 111~\text{MeV}$.

Besides,  \textbf{Table \ref{tab:width2}} presents our CGAN predictions for the decay widths of certain mesons, 
for which experimental values are not available, alongside the previously reported DNN estimates.
These results may assist experimental groups in their search for the corresponding resonances and in determining their decay widths.

When comparing the predicted results of our CGAN model with those from previous DNN models 
and available experimental data for the masses and widths of various mesons, 
it is clear that the CGAN model outperforms the DNN approach. 
The CGAN framework consistently shows a smaller discrepancy between the experimental and predicted values, 
indicating a higher degree of accuracy in its predictions.

\begin{table}[h!]
\begin{center}
\renewcommand{\arraystretch}{1.4}
\scalebox{0.68}{
\begin{tabular}{|c|c|c|c|c|}\cline{1-5}
Meson & $I\,(J^{PC})$	& Width (MeV)& DNN& CGAN \!\!\! \\ \hline
\hline
$a_0(980)$	 & $1^-\,(0^{++})$	&  $97 \pm 1.9 \pm 5.7$ \cite{CrystalBarrel:2019zqh} & $113\pm28$ & $101\pm 33 $\\ \hline
$a_0(980)_{exotic}$	 & $1^-\,(0^{++})$	& $97 \pm 1.9 \pm 5.7$ \cite{CrystalBarrel:2019zqh} & $179\pm 84$& $210\pm 80 $\\ \hline

$f_0(980) $	 & $0^+\,(0^{++})$	&$ 10-100 $ \cite{ParticleDataGroup:2022pth} & $105\pm34 $& $82\pm 32 $\\ \hline
$f_0(980)_{exotic} $	 & $0^+\,(0^{++})$	&$ 10-100 $\cite{ParticleDataGroup:2022pth}& $120\pm 58$& $105\pm80 $\\ \hline
$ f_0(1370)$	 & $0^+\,(0^{++})$	& $200 - 500$ \cite{ParticleDataGroup:2022pth}  & $107\pm 40$& $197\pm 62 $\\ \hline
$ D^*(2007)^0$	 & $1/2\,(1^-)$	& $< 2.1$ ($CL = 90 \%$) \cite{Abachi:1988fw}  & $4.6\pm1.2 $& $1.6\pm 0.8 $\\ \hline
$ D_{s_0}^{*}(2317)^{\pm}$	 & $0\,(0^+)$	& $ <3.8$ ($CL = 95 \%$) \cite{BaBar:2006eep} & $3.1\pm 1.7 $& $3.2\pm 0.5 $\\ \hline
$D_{s_0}^{*}(2317)^\pm_{exotic} $	 & $0\,(0^+)$	& $ <3.8$ ($CL = 95 \%$) \cite{BaBar:2006eep} & $47\pm 23 $& $11.7\pm7 $\\ \hline
$ D_{s1}(2460)^{\pm}$	 & $0\,(1^+)$	& $< 3.5$ ($CL = 95 \%$) \cite{BaBar:2006eep}  & $3.4\pm1.6 $& $3.2\pm 1.2 $\\ \hline
$ D_{s1}(2460)^\pm_{exotic} $	 & $0\,(1^+)$	& $< 3.5$ ($CL = 95 \%$) \cite{BaBar:2006eep}  & $31\pm 21 $& $8.2\pm 3 $\\ \hline
$\psi_2(3823) $	 & $0\,(2^{--})$	& $ < 2.9$ ($CL = 90 \%$) \cite{BESIII:2022yga} & $6\pm4 $& $2.7\pm 1.5$\\ \hline
$\eta_b(2s) $	 & $0^+\,(0^{-+})$	& $< 24 (CL = 90 \%)$ \cite{Belle:2012fkf}& $54\pm24 $& $12.8\pm 4$\\ \hline
$ K_0^*(700)$	 & $1/2\,(0^+)$	& $ 468 \pm 30$ \cite{ParticleDataGroup:2022pth}& $328\pm 140$& $461\pm 146 $\\ \hline
$K_0^{*}(1430) $	 & $1/2\,(0^+)$	& $270 \pm 80 $ \cite{ParticleDataGroup:2022pth}  & $254\pm 84 $& $274\pm 111 $\\ \hline

\end{tabular}
}
	\caption{Predictions of our CGAN model for the width of some exotic mesons (in units of MeV). 
Results are presented alongside some experimental data and our previous DNN model predictions in Ref. \cite{Malekhosseini:2024eot}.}\label{tab:width1}
\end{center}
\end{table}

\begin{table}[h!]
\begin{center}
\renewcommand{\arraystretch}{1.4}
\scalebox{0.68}{
\begin{tabular}{|c|c|c|c|}\cline{1-4}
Meson & $I\,(J^{PC})$	& Width (DNN) & Width (CGAN) \!\!\! \\ \hline
\hline
$B^* $	 & $1/2\,(1^-)$	& $0.84\pm0.56 $ &  $0.02\pm 0.01 $\\ \hline
$B^*_{s_0} $	 & $0\,(1^-)$	&  $0.63\pm 0.2 $ &  $0.04\pm 0.01 $\\ \hline
$B_c(2s)^{\pm} $	 & $0\,(0^-)$	& $2.4\pm 1 $ &  $5.9 \pm 3.1 $\\ \hline
$ \chi_{b_0}(1p)$	 & $0^+\,(0^{++})$	 & $45\pm 20 $ &  $3.6\pm 1.7 $\\ \hline
$\chi_{b_0}(2p) $	 & $0^+\,(0^{++})$	& $50\pm 28$ &  $4.5\pm 1.9  $\\ \hline
$ \chi_{b_1}(1p)$	 & $0^+\,(1^{++})$	 & $16\pm9 $ &  $2.2\pm1.5  $\\ \hline
$ \chi_{b_1}(2p)$	 & $0^+\,(1^{++})$	 & $20\pm 10$ &  $3.7\pm0.7  $\\ \hline
$\chi_{b_1}(3p)$	 & $0^+\,(1^{++})$& $29\pm19 $ &  $5.3\pm 1.2 $\\ \hline
$ \chi_{b_2}(1p)$	 & $0^+\,(2^{++})$	& $21\pm7 $ &  $2.8\pm 1.6  $\\ \hline
$\chi_{b_2}(2p)$	 & $0^+\,(2^{++})$	& $23\pm 11$ &  $ 4.1\pm 2.1 $\\ \hline
$ \chi_{b_2}(3p)$	 & $0^+\,(2^{++})$ & $39\pm 19$ &  $5.48\pm 2.1 $\\ \hline
$ K_0,\bar{K}_0$	 & $1/2\,(0^-)$	&   $(3.34\pm 1.8)\times 10^{-6} $ &  $(1.2\pm 0.7)\times 10^{-6}  $\\ \hline

\end{tabular}
}
	\caption{Predictions of our CGAN model for the width of some exotic mesons (in units of MeV). 
Results are presented alongside our previous DNN model predictions in Ref. \cite{Malekhosseini:2024eot}.}\label{tab:width2}
\end{center}
\end{table}

\section{Summary and conclusions}\label{SC}
CGAN frameworks are a powerful class of ML models, capable of generating data conditioned on a specific input or label. 
In contrast to standard GANs, CGANs can generate more targeted outputs by conditioning the model on additional information, 
such as particle properties or experimental conditions.
Furthermore, when the input data is limited or hard to obtain, 
CGANs can generate additional synthetic data, 
enhancing training datasets for ML models. In this study, for the first time,
we applied the CGAN framework to augment the mesonic data, 
preserving the inherent characteristics of the original dataset.
We then employed the CGANs to predict the mass and width of both ordinary and exotic mesons 
based on their flavor content and corresponding quantum numbers.
Combination of the augmented training data and the inherent advantages 
of the CGAN architecture can lead to predictions 
with smaller uncertainties and better alignment with experimental results. 
We present the numerical results from our CGAN model for the mesons' mass and decay width,
compared to the corresponding experimental values and our previous DNN predictions.
The CGAN model offers a significant improvement over the DNN model in predicting the mass and decay width of various mesons. 
The more consistent predictions from the CGAN model highlight its potential as a more effective tool for making 
reliable predictions in the study of meson properties, including both ordinary and exotic meson configurations. 
These improvements suggest that the GAN model is a promising approach for exploring the internal structures of mesons, 
such as the possibility of tetraquark states.
In contrast, the DNN model, without the benefit of augmented training data, struggles to achieve the same level of accuracy. 
It is important to clarify that the goal of this work is not to replace or redefine existing theoretical models of hadron structure, but rather to complement them with a data-driven predictive framework. By employing CGANs trained on fundamental meson features, we aim to provide practical estimations of mass and width for mesons whose experimental measurements are incomplete or unavailable. This predictive capability, grounded in essential quantum numbers, can be valuable for guiding future experimental searches and for supporting phenomenological studies in the field. We recognize that further theoretical insight into hadron structure requires more comprehensive approaches, which could be combined with or built upon our framework in future research.
A next prominent step will be to explore key features of the baryons, pentaquarks
and possible molecular dibaryons through CGAN techniques.
Also, CGANs can be used to simulate particle collision events. 
This could be particularly useful in situations where 
traditional simulations are computationally expensive or slow.
Given initial conditions (e.g., particle type, momentum), 
CGANs can be used to generate predictions about possible decay modes or 
interactions between particles, which will be valuable for understanding rare processes.
In conclusion, the application of CGANs provides a promising 
approach to enhancing the power of predictive tools in particle physics.

\section*{ACKNOWLEDGEMENTS}
S.~R. and M.~M. would like to express their heartfelt gratitude to the organizers of the MITP Summer School on "Machine Learning in Particle Theory" for their invaluable support, insightful lectures, and the opportunity to engage with cutting-edge advancements in the intersection of machine learning and particle physics. S.~R., M.~M., and K.~A are grateful to the CERN-TH division for their warm hospitality.

\end{document}